\RequirePackage{fix-cm}
\documentclass[smallextended]{svjour3}       
\smartqed  
\usepackage{graphicx}
\usepackage{algorithmic,algorithm}
\usepackage{color,colortbl}
\usepackage[colorlinks,linkcolor=black]{hyperref}
\usepackage{amsmath,amsfonts,amssymb,graphicx}
\usepackage{latexsym}
\newcommand{\bra}[1]{\langle{#1}|} 
\newcommand{\ket}[1]{|{#1}\rangle} 
 \journalname{Int J Theor Phys}
\begin{document}

\title{Quantum associative memory with improved distributed queries
}

\titlerunning{Improved quantum associative memory}  

\author{J.-P. TCHAPET NJAFA \and S.G. NANA ENGO \and Paul WOAFO}

\authorrunning{J.-P. TCHAPET NJAFA et al.} 

\institute{J.-P. TCHAPET NJAFA \and S.G. NANA ENGO \at
              Department of Physics, University of Ngaoundere, POB 454
Ngaoundere, Cameroon\\
              \\
              \email{nanaengo@gmail.com}
           \and
           Paul WOAFO \at
              Laboratory of Modelling and Simulation in Engineering and
Biological Physics, Department of Physics, Faculty of Science, University of
Yaounde I, POB 814 Yaounde, Cameroon
}

\date{Received: date / Accepted: date}

\maketitle

\begin{abstract}
The paper proposes an improved quantum associative algorithm with
distributed query based on model proposed by Ezhov \emph{et al.}. We introduce
two modifications of the query that optimized data retrieval of correct
multi-patterns simultaneously for any rate of the number of the recognition
pattern on the total patterns. Simulation results are given.

\keywords{Hopfield model \and quantum associative memory \and pattern
recognition \and Grover's algorithm \and distributed queries}
\end{abstract}

\section{Introduction}
\label{intro}

Since the last two decades, there is a growing interest in quantum computing,
due to the improvement in memory size (such as super dense coding) and the
speed-up in computing time. Three advantages of quantum computing make it
possible: (i) the \emph{quantum parallelism} which is expressed in the principle
of superposition and provides an advantage in processing huge data sets; (ii)
the \emph{entanglement}, the strange quantum phenomenon that links qubits across
distances and provides the possibility of measuring the state of all qubits in a
register whose values are interdependent; (iii) the \emph{unitary} of quantum
gates which ensure reversibility and therefore overcome energy dissipation.


Associative memory is an important tool for intelligent control, artificial
intelligence and pattern recognition. Considering that quantum information is
information stored as a property of a quantum system e.g., the polarization of a
photon, or the spin of an electron, several approaches of quantum associative
memory for pattern recognition have been proposed.

Perus \emph{et al.} worked with quantum-wave Hopfield-like algorithm that has
been successfully tested in computer simulations of concrete pattern recognition
applications \cite{perus-bischof-03,PerusBischof:2003,perus.2004-43}.

The model proposed by Trugenberger makes a link between Artificial Neural
Network (ANN)-like and logic-gate-based branches of quantum pattern recognition.
The model is related to the fact that a special case of Hebbian memory-storage
is equivalent to quantum-implementable \texttt{NOT XOR} gate
\cite{Trugenberger:02}. However, some critical view had been added indicating
that the advantage of quantum states as memories for classical information is
not clearly demonstrated.

Ventura and Martinez have built a model of quantum associative memory where the
stored patterns are considered to be basis states of the memory quantum state
(\emph {inclusive} method of quantum superposition). Unitary rotation operators
increase the probability to recall the basis state associated with the input
pattern \cite{ventura.is00,ventura.ijcnn01,ventura.jcis02}. The retrieval
algorithm is based on the well known Grover's quantum search algorithm in an
unsorted database, which is also an amplitude amplification of the desired state
\cite{grover.1998-80}. In order to overcome the limitation of that model to only
solve the completion problem by doing data retrieving from noisy data, Ezhov et
\emph{al.} have used an \emph {exclusive} method of quantum superposition and
Grover's algorithm with distributed query \cite{ezhov.is00}. However, their
model still produces non-negligible probability of irrelevant classification.

Recently, Zhou and Ding have presented a new quantum multi-pattern recognition
method, based on the improved Grover's algorithm proposed by Li and Li
\cite{li2007phase}, which can recognize multi-pattern simultaneously with the
probability of $100\%$ \cite{ZhouDing:2008}. The method introduces a new design
scheme of initializing quantum state and quantum encoding on the pattern set.
However, there is an important constraint that the rate of the number of the
recognized pattern on the total patterns should be over $\frac{1}{3}$.

This paper suggests, through an \emph{oracle} operator $\mathcal{I}_{M}$, two
modifications of the query that can optimize data retrieval of correct
multi-patterns simultaneously in the Ezhov's model of quantum associative memory
without any constraint. In the first modification, $\mathcal{I}_{M}$ invert only
the probability amplitudes of the memory patterns states as in the Ventura's
model. In the second modification, $\mathcal{I}_{M}$ invert the probability
amplitudes of all the states over the average amplitude of the query state
centered on the $m$ patterns of the learning.

The main steps of the approach of these models are \cite{ventura.jcis02}:
\begin{itemize}
 \item Construction of a quantum classification system that approximates the
function from which set of patterns $M$ (\emph{memory states}) was drawn;

  \item Classification of instances using appropriated Grover's quantum search
algorithm (the time evolution step);

  \item Observation of the system.
\end{itemize}

The paper is organized as follows: Section \ref{sec:Vent} briefly present basic
ideas of Ventura's and Ezhov's models respectively. Section \ref{sec:modif} is
used to introduce our approach and show the comparing results with the above
mentioned models. In Section \ref{esc:concl} we summarize the paper.

\section{Ventura's model and Ezhov's model}
\label{sec:Vent}

The main purpose of the quantum associative memory build by Ventura and
Martinez is \emph{pattern completion}  \cite{ventura.is00}. That is, it can
restore the full pattern from partial, but exact, part one. The memory use a
storage algorithm and Grover's quantum search algorithm for retrieving the
patterns.

Grover's quantum search algorithm can be considered as a rotation of the state
vectors in two-dimensional Hilbert space generated by the initial and target
vectors\cite{grover.1998-80}. The amplitude of the target state increases
towards its maximum while the amplitudes of other states decreases after a
certain number of iterations.

Each neuron of the memory is a qubit that can be in a state $\ket{0}$, $\ket{1}$
or a superposed state $\alpha\ket{0}+\beta\ket{1}$, with $|\alpha|^2+|\beta|^2
=1$. $|\alpha|^2$ and $|\beta|^2$ are respectively the probability of state
$\ket{0}$ and state $\ket{1}$. As with a register of $n$ qubits, one can compute
at the same time all the $2^n$ numbers by using the following superposition
state
\begin{equation}
 \ket{\psi}=\sum_{x=0}^{2^n-1}c_x\ket{x},\,\sum_{x=0}^{2^n-1}|c_x|^2=1,
 \label{eq:sup0}
\end{equation}
a quantum associative memory can learn or store $2^n$ patterns. In the standard
Grover's algorithm, Eq. (\ref{eq:sup0}) is obtained by applying $n$ times the
Walsh-Hadamard gate
\begin{equation}
\mathtt{W}=\frac{1}{\sqrt{2}}(\ket{0}\bra{k}+(-1)^{k}\ket{1}\bra{k}),\
k=\{0,1\},
\end{equation}
to the initial state $\ket{0}$.

\subsection{Storage algorithm}

To generate a quantum register (\ref{eq:sup0}) in the superposition of only
desired states from an initial state of $n$ qubits of the network (\emph
{inclusive} method), Ventura and Martinez used the storage algorithm that they
named algorithm of \textbf{initializing the amplitude distribution of a quantum
state}. It works in a polynomial time and separate the initial state into the
already stored patterns term and ready to process a new pattern term. The main
operator of this algorithm is the 2-qubit controlled gate \emph{state
generation} \cite{ventura.fopl99,Trugenberger:02},
\begin{equation}
\begin{split}
  \texttt{CS}^p & =\ket{0}\bra{0}\otimes\mathbb{I}+\ket{1}\bra{1}\otimes
\texttt{S}^p  =\mathtt{diag}(\mathbb{I},\texttt{S}^p),\\
  \texttt{S}^p & =\begin{pmatrix}
         \sqrt{\frac{p-1}{p}} & -\frac{1}{\sqrt{p}}\\
          \frac{1}{\sqrt{p}} & \sqrt{\frac{p-1}{p}}
	\end{pmatrix},
\end{split}
\label{eq:CSp}
\end{equation}
for $m\geq p\geq 1$, where $m\leq 2^{n}$ the number of pattern of length $n$ to
be store\footnote{Generally $m\ll2^{n}$.}, each $p$ is associate to a pattern.
$\mathbb{I}$ denotes the two-dimensional identity matrix. The operator
(\ref{eq:CSp}) separates out the new pattern to be store by assigning to it
small amplitude so that others operators can't act on it. To do that, we use
three registers of $n$, $n-1$ and $2$-qubits:
\begin{itemize}
\item $\ket{x}=\ket{x_1\dots x_n}$ the register where the $m$ patterns of length
$n$ will be stored;

\item $\ket{g}=\ket{g_1\dots g_{n-1}}$ a register used like workspace to
identify and mark a particular state;

\item $\ket{c}=\ket{c_1c_2}$ a register of two-qubits of control, that is the
 operator $\mathtt{CS}^{p}$ acts when $\ket{c_1}=\ket{1}$.
\end{itemize}
At the end of the algorithm there is no entanglement between the $x$-register
and the two others which are respectively at $\ket{0^{\otimes n-1}}\equiv
\ket{0_1\dots0_{n-1}}$ and $\ket{0^{\otimes 2}}\equiv\ket{00}$.

The simplified form of the storage algorithm is:
\begin{algorithm}
\caption{Simplified form of algorithm \ref{alg:apprend}
(Ref.~\protect\cite{ventura.fopl99})}
\label{alg:resumapprend}
\begin{algorithmic}[1]
\STATE $\ket{\psi}=\ket{x_{1}\dots x_{n},g_{1}\dots
g_{n-1},c_{1}c_{2}}\equiv\ket{0_1\dots0_n,0_1\dots0_{n-1},00}$;
\COMMENT{Initialize the register}
\FOR{ $m\geq p\geq 1$}
\STATE $\ket{\psi}=\mathtt{FLIP}\ket{\psi}$;
\COMMENT{Generate the state}
\STATE $\ket{\psi}=\mathtt{CS}^{p}\ket{\psi}$;
\COMMENT{Apply $\mathtt{CS}^{p}$ operator}
\STATE $\ket{\psi}=\mathtt{SAVE}\ket{\psi}$;
\COMMENT{Save the state}
\ENDFOR
\STATE $\ket{\psi}=\mathtt{NOT}_{c_{2}}\ket{\psi}$;
\STATE Observe the system.
\end{algorithmic}
\end{algorithm}

The $\mathtt{FLIP}$ operator,
\begin{equation}
 \mathtt{FLIP}=\mathtt{CNOT}_{(c_{2}c_{1})}^{0}\mathtt{CNOT}_{(c_{2}x_{j})}^{0},
\,(1\leq j\leq n,\,z_{pj}\neq z_{(p+1)j}),
\end{equation}
change the qubits state of the $x$-register when $\ket{c_1}=\ket{0}$ so that
they correspond to the states $\ket{P}$ associated to patterns. The
$\mathtt{SAVE}$ operator makes the state with the smaller amplitude a permanent
representation of the pattern being processed and resets the other to generate a
new state for the new pattern. At the end of the whole process the system is in
the state
\begin{equation}
\ket{\psi}=\frac{1}{\sqrt{m}}\sum^{m}_{1}\ket{P}.
\label{eq:psi}
\end{equation}
called \emph{blank memory} in the sense that all possible states have the same
probability of being recovered upon measurement \cite{Trugenberger:02}. The
number of steps of the storage algorithm is $\mathcal{O}(mn)$ which is optimal
because reading each instance once cannot be done faster than that.

\subsection{Retrieving algorithm}

The associative memory proposed by Ventura and Martinez uses for retrieving
information a modified version of \textbf{Grover's} search algorithm of an
unsorted database. The original Grover's algorithm has been modified in order to
include cases where not all possible pattern are represented and where more than
one target state is to be found. The reader is referred to \cite{ventura.is00}
for more details. It should be noted that Grover's algorithm use only
$\mathcal{O}(\sqrt{\frac{N}{m}})$ steps to retrieve $m$ elements in disordered
list of $N=2^n$ elements, while in classics algorithms the best use
$\mathcal{O}(\frac{N}{m})$ steps.

\subsection{Ezhov's model}
\label{sec:Ezh}

As mentioned in the last section, the associative memory proposed by Ventura and
Martinez can only do completion data. That is bits sequence shown to the network
should be identical to a part of bits of one of memorized patterns. In order to
overcome this limitation, Ezhov et \emph{al.} have introduced a metric into the
quantum search algorithm in the form of distributed queries \cite{ezhov.is00}.
The model is able to retrieve memory states with probability proportional to the
amplitudes these states have in the query. Their quantum memory can retrieve
valid stored patterns from a noisy data.

The model use the \emph{exclusion learning approach} in which the system is in
superposition of all the possible states, except the patterns states. If $M$ is
the set of patterns and $m$ the number of patterns of length $n$,
\begin{equation}
 \ket{\Psi}=\frac{1}{\sqrt{N-m}}\sum^{N-1}_{x\notin M}\ket{x},\,N=2^n.
\label{eq:Psi}
\end{equation}
In other words, the exclusion approach for the learning pattern included each
point not in $M$ with nonzero coefficient while those points in $M$ have zero
coefficients.

\begin{figure}[!htb]
 \centering
 \includegraphics[scale=0.18]{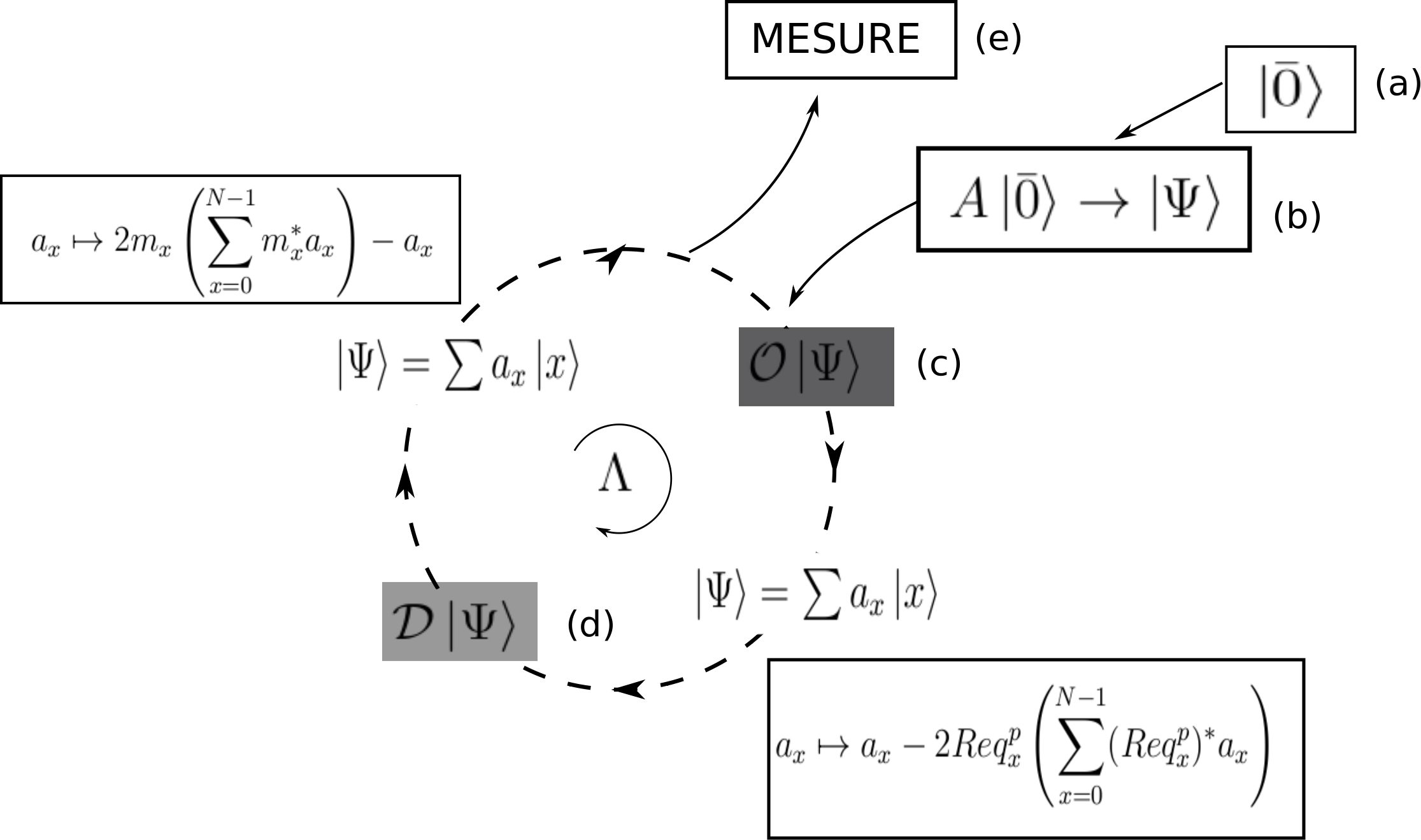}%
 \vspace*{13pt}
\caption{The Brickman's diagram \protect\cite{brickman.2005} of algorithm
\ref{alg:requet}. (a) Qubits are initialize to the state $\ket{\bar{0}}$. (b)
The operator $A$ performs the learning of the set $M$ using the
\emph{exclusion approach} and gives to all the possibles states the same
probability amplitude $a_x$. (c) The operator oracle $\mathcal{O}$ inverts
probability amplitude of all the states over the average amplitude of the query
state by changing $a_x$ to $a_{x}-2Req^{p}_{x}\left(\sum^{N-1}_{x=0}
(Req^{p}_{x})^{*}a_{x}\right)$. (d) The operator diffusion $\mathcal{D}$ inverts
the probability amplitude of the states of $\ket{\Psi}$ over their average
amplitude and for the others over the value $0$ by changing the amplitudes of
probability $a_x$ to $2m_{x}\left(\sum^{N-1}_{x=0}m^{*}_{x}a_{x}\right)-a_{x}$.
(e) After $\Lambda$ applications of steps (c) and (d) the system is observed.}
 \label{fig:cicle}
\end{figure}

The distributed query is in the following superposed states
\begin{equation}
 \ket{Req^{p}}=\sum^{N-1}_{x=0}Req^{p}_{x}\ket{x},
\end{equation}
where $Req^{p}_{x}$ obey to binomial distribution
\begin{equation}
 \|Req^{p}_x\|^{2}=a^{d_H(p,x)}(1-a)^{n-d_H(p,x)}.
\label{eq:re}
\end{equation}
In equation (\ref{eq:re})
\begin{itemize}
 \item $p$ marks the state $\ket{p}$ which is referred as the query center;
\item $0<a<\frac{1}{2}$ is an arbitrary value that regulates the width of the
distribution;
\item the \textbf{Hamming distance} $d_H(p,x)=|p-x|$ between $\ket{x}$ and
$\ket{p}$ is an important tool which gives the correlation between input and
output;
\item the amplitudes are such that $\sum_{x}\|Req^{p}_x\|^{2}=1$.
\end{itemize}
The corresponding memory's algorithm is give by Algorithm \ref{alg:requet} and
the associate Brickman's diagram \cite{brickman.2005} by Figure \ref{fig:cicle}.

\begin{algorithm}
\caption{Quantum associative memory with distributed queries
(Ref.~\protect\cite{ezhov.is00})}
\label{alg:requet}
\begin{algorithmic}[1]
\STATE $\ket{0_10_2\dots 0_n}\equiv \ket{\bar{0}}$;
\COMMENT{Initialize the register}
\STATE $\ket{\Psi}=A\ket{\bar{0}}=\frac{1}{\sqrt{N-m}}\sum^{N-1}_{x\notin
M}\ket{x}$;
\COMMENT{Learn the patterns using exclusion approach}
\REPEAT
\STATE \label{etap4} Apply the operator oracle $\mathcal{O}$ to the register;
\STATE \label{etap5} Apply the operator diffusion $\mathcal{D}$ to the register;
\STATE $i=i+1$;
\UNTIL{$i>\Lambda$}
\STATE Observe the system.
\end{algorithmic}
\end{algorithm}

In the Algorithm \ref{alg:requet} or in the associate Brickman's diagram of
Figure \ref{fig:cicle},
\begin{itemize}
 \item $\mathcal{O}$ is the operator oracle which invert the phase of the
query state $\ket{Req^{p}}$,
 \begin{align}
 \mathcal{O}& =\mathbb{I}-(1-e^{i\pi})\ket{Req^{p}}\bra{Req^{p}},\\
  \mathcal{O} &:a_{x}\mapsto a_{x}-2Req^{p}_{x}\left(\sum^{2^{n}-1}_{x=0}
(Req^{p}_{x})^{*}a_{x}\right),
\label{eq:O}
\end{align}
where $a_{x}$ is the probability amplitude of the state $\ket{x}$.
\item $\mathcal{D}$ is the operator diffusion which invert the probability
amplitude of the states of $\ket{\Psi}$ over their average amplitude and for the
others over the value $0$.
 \begin{align}
  \mathcal{D}& =(1-e^{i\pi})\ket{\Psi}\bra{\Psi}-\mathbb{I},\\
 \mathcal{D} &:a_{x}\mapsto 2m_{x}\left(\sum^{N-1}_{x=0}m^{*}_{x}a_{x}\right)
-a_{x}.
\label{eq:D}
\end{align}
where $m_{x}$ is the probability amplitude of a state of
$\ket{\Psi}$.
\item $\Lambda$ is the number of iterations that whilst the maximal value of
amplitudes, which must be as far as possible nearest to an integer,
\begin{equation}
 \Lambda=T(\frac{1}{4}+\alpha),\,T=\frac{2\pi}{\omega},\,\alpha\in\mathbb{N},
\label{eq:lambda}
\end{equation}
with the Grover's frequency
\begin{equation}
\label{equaB}
 \omega=2\arcsin{B},
B=\frac{1}{\sqrt{N-m}}\sum^{N-1}_{x=0,x\notin M}Req^{p}_{x}.
\end{equation}
\end{itemize}

\begin{example}
\label{exa:exam1}
In order to help clarify, consider a 3-qubits memory where the patterns for the
learning are $\ket{010}=\ket{2}$ and $\ket{100}=\ket{4}$ and the distributed
query centered on $\ket{011}=\ket{3}$. For $a=\frac{1}{4}$,
\begin{align}
\ket{Req^{3}} & =\frac{\sqrt{3}}{8}\ket{0}+\frac{3}{8}\ket{1}+\frac{3}{8}\ket{2}
+\frac{3\sqrt{3}}{8} \ket{3}+\frac{1}{8}\ket{4}+\frac{\sqrt{3}}{8}\ket{5}
+\frac{\sqrt{3}}{8}\ket{6} +\frac{3} {8}\ket{7},\\
 B & =\frac{1}{\sqrt{2^{3}-2}}\sum^{7}_{x=0,x\neq 2,x\neq 4}Req^{3}_{x}
=\frac{6+6\sqrt{3}}{ 8\sqrt {6}},\\
 \omega &=2\arcsin{\frac{6+6\sqrt{3}}{8\sqrt{6}}}=0.63\pi\,\Leftrightarrow
T=3.17\Longrightarrow\text{ for }\alpha=1,\,\Lambda=4.
\end{align}
The steps \ref{etap4} and \ref{etap5} of the Algorithm \ref{alg:requet} will be
repeated $4$ times.

The operator oracle (\ref{eq:O}) is
\begin{align}
\begin{split}
 \mathcal{O} &=\mathbb{I}-\frac{1}{32}\left(\sqrt{3}\ket{0}+3
\ket{1}+3\ket{2}+3\sqrt{3}\ket{3}+\ket{4}+\sqrt{3}\ket{5}
+\sqrt{3}\ket{6}+3\ket {7}\right)\\ & \left(\sqrt{3}\bra{0}+3
\bra{1}+3\bra{2}+3\sqrt{3}\bra{3}+\bra{4}+\sqrt{3}\bra{5}
+\sqrt{3}\bra{6}+3\bra {7}\right).
\end{split}
\end{align}

As after the learning process the quantum memory is in the state
\begin{equation}
\ket{\Psi}=\frac{1}{\sqrt{6}}(\ket{0}+\ket{1}+\ket{3}+\ket{5}+\ket {6}+\ket{7}),
\end{equation}
the operator diffusion (\ref{eq:D}) is
\begin{equation}
 \mathcal{D}=\frac{1}{3}\left(\ket{0}+\ket{1}+\ket{3}+\ket{5}+\ket{6}+\ket{7
}\right)\left(\bra{0}+\bra{1}+\bra{3}+\bra{5}+\bra{6}+\bra{7}\right)-\mathbb{I}.
\end{equation}

At the end of the $4$ iterations the register is in the state
\begin{equation}
\begin{split}
\ket{\Psi^{4}}=-0.257\ket{0}+0.031\ket{1}+0.683\ket{2}+0.531\ket{3}
+0.228\ket{4}\\-0.257\ket{5}-0.257\ket{6}+0.031\ket{7}.
\end{split}
\end{equation}
The probability to retrieve the memory states, $\ket{2}$ and $\ket{4}$, is
$0.683^2+0.228^2=51.85\%$. As expected, the memory state closest in Hamming
distance to the query center state, $\ket{2}$, presented the best probability
($46.65\%$).
\end{example}
\begin{figure}[ptbh]
\centering
\begin{minipage}[c]{.55\linewidth}
\centering
 \includegraphics[scale=.6,keepaspectratio=true]{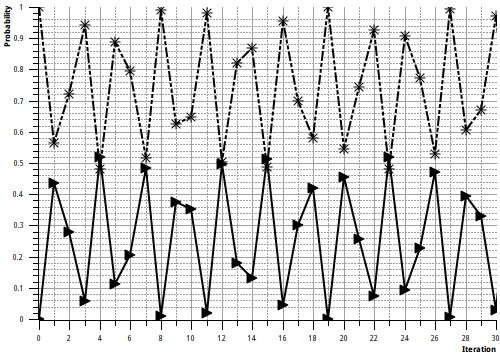}
\vspace*{13pt}
 \caption{Probability evolution with the number of iterations in Ezhov's model
for 3-qubits memory patterns. The solid line represents the probability $P_c$ of
a correct recognition and the dotted line the probability $P_w$ of an incorrect
recognition.}
 \label{fig:EvolEzhov}
\end{minipage} \hfill
\begin{minipage}[c]{.32\linewidth}
\centering
 \includegraphics[scale=.6]{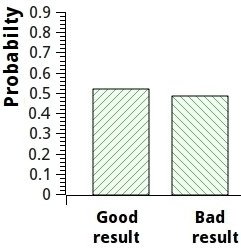}
 \vspace*{13pt}
\caption{Probability of correct recognition for a set of two example patterns
in Ezhov's model.}
 \label{fig:ProbaEzhov}
\end{minipage}
\end{figure}
Figure \ref{fig:EvolEzhov} shows the probability of observing the correct
recognition upon system measurement versus Grover's search iterations. The solid
line represents the probability $P_c$ of a correct recognition and the dotted
line the probability $P_w$ of an incorrect recognition. Note that the periodic
nature of the algorithm clearly appears, and it can be seen that the probability
of success $P_c$ is maximized after four iterations. At $\Lambda=4$, the ratio
\begin{equation}
\frac{P_c}{P_w}=1.08,
\label{eq:ratioEzhov}
\end{equation}
that could be considered as the recognition efficiency of the memory patterns,
shows that the confidence that the recognition is correct fair in the Ezhov's
model. This clearly appears in the Figure \ref{fig:ProbaEzhov} that gives the
graphic representation of the probabilities of correct and bad recognition of
this example.

It should be pointed out that if the approximation
\begin{equation}
 B\simeq\frac{1}{\sqrt{N}}\sum^{N-1}_{x=0}Req^{p}_{x},\, N\gg m,
\end{equation}
is wrongly use in Example \ref{exa:exam1}, the number of iterations increase to
$\Lambda=9$ and the probability to retrieve the memory states $\ket{2}$ and
$\ket{4}$ is reduce to $36.00\%$ and therefore the recognition efficiency is
\begin{equation}
\frac{P_c}{P_w}=0.56.
\end{equation}

\section{Improved quantum associative algorithm with distributed query}
\label{sec:modif}

\begin{figure}[!ht]
 \centering
 \includegraphics[scale=0.15]{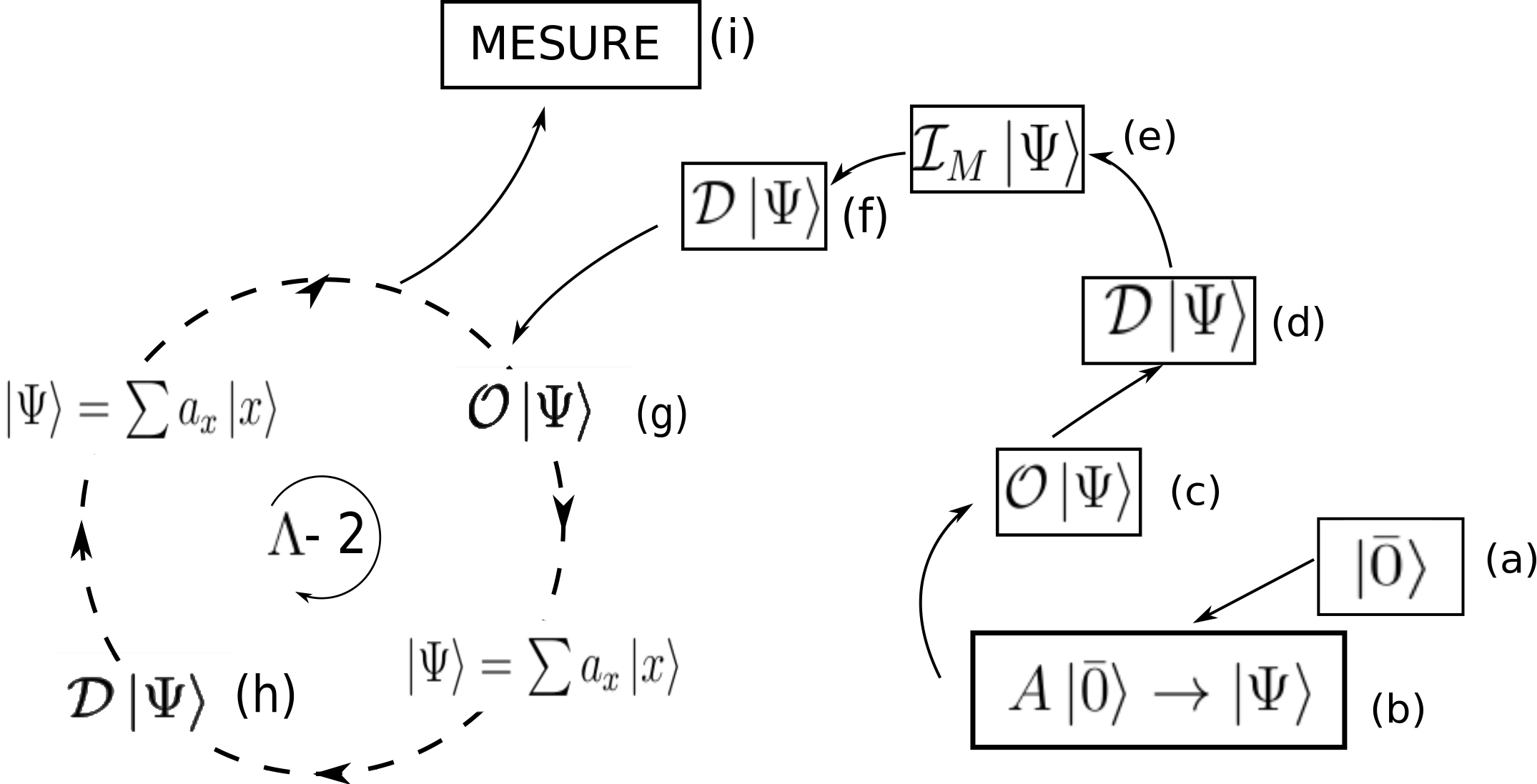}%
\vspace*{13pt}
\caption{Brickman's diagram \protect\cite{brickman.2005} of Algorithm
\ref{alg:requetm}. The steps here are the same as in the diagram of Figure
\ref{fig:cicle}, with however two modifications: (e) The new operator
$\mathcal{I}_{M}$ acts as the operator oracle and (f) the operator
diffusion $\mathcal{D}$ is apply again before the $\Lambda-2$ iterations.}
 \label{fig:ciclem}
\end{figure}

In order to improve the quantum associative memory with distributed query such
that it optimize the probability of retrieving the learned patterns, even for
the biggest Hamming distance from the query center, we proposed the Algorithm
\ref{alg:requetm} illustrate by the Brickman's diagram of Figure
\ref{fig:ciclem}, with an operator $\mathcal{I}_{M}$.
\begin{algorithm}
\caption{Improve quantum associative memory with distributed query}
\label{alg:requetm}
\begin{algorithmic}[1]
\STATE $\ket{0_10_2\dots 0_n}\equiv \ket{\bar{0}}$;
\COMMENT{Initialize the register}
\STATE $\ket{\Psi}=A\ket{\bar{0}}=\frac{1}{\sqrt{N-m}}\sum^{N-1}_{x\notin
M}\ket{x}$;
\COMMENT{Learn the patterns using exclusion approach}
\STATE \label{etapp3} Apply the operator oracle $\mathcal{O}$ to the register;
\STATE Apply the operator diffusion $\mathcal{D}$ to the register;
\STATE \label{etapp5} Apply operator $\mathcal{I}_{M}$ to the register;
\STATE \label{etapp6} Apply the operator diffusion $\mathcal{D}$ to the
register;
\REPEAT
\STATE \label{etap6} Apply the operator oracle $\mathcal{O}$ to the register;
\STATE \label{etap7} Apply the operator diffusion $\mathcal{D}$ to the register;
\STATE $i=i+1$;
\UNTIL{$i>\Lambda-2$}
\STATE Observe the system.
\end{algorithmic}
\end{algorithm}

Two cases we be will considered for the operator $\mathcal{I}_{M}$:
\begin{itemize}
\item[\textbf{C1}:] $\mathcal{I}_{M}$ invert only the phase of the memory
patterns states as in the Ventura's model,
\begin{align}
 \mathcal{I}_{M} &=\mathbb{I}-(1-e^{i\pi})\ket{\varphi}\bra{\varphi},\,
\ket{\varphi}\bra{\varphi}=\sum_{x\in M}\ket{x}\bra{x},\\
 \mathcal{I}_{M} & :a_{x}\mapsto
\begin{cases}
-a_{x}\text{ if }\ket{x}\in M\\
a_{x}\text{ if not.}
\end{cases}
\end{align}
$\forall x\in M$, the grover operator act as
\begin{align}
 \begin{split}
\mathcal{D}\mathcal{I}_M\ket{\varphi}&=(2\ket{\Psi}\bra{\Psi}-\mathbb{I}
+2\ket{\varphi}\bra{\varphi})\ket{\varphi}\\
& =2\ket{\Psi}\langle\Psi\ket{\varphi}-\ket{\varphi}+2\ket{\varphi}
\langle\varphi\ket{\varphi}=\ket{\varphi}.
 \end{split}
\end{align}

 \item[\textbf{C2}:] $\mathcal{I}_{M}$ is formally identical to the operator
oracle $\mathcal{O}$ of Eq. (\ref{eq:O}),
\begin{equation}
  \mathcal{I}_{M} :a_{x}\mapsto a_{x}-2REQ_{x}\left(\sum^{N-1}_{x=0}
(REQ_{x})^{*}a_{x}\right),
\end{equation}
with
\begin{equation}
 \label{eq:RE}
\|REQ_x\|^{2} =\frac{1}{k}\sum_{p}a_{b}^{d_H(b,x)}(1-a_{b})^{n-d_H(b,x)},
\end{equation}
where we consider that the distribution have $k$ centers and $0<a_b<\frac{1}{2}$
is an arbitrary value that regulates the width distribution around the center
$b$. But in this paper, we will consider that
\begin{equation}
 \label{eq:REm}
\|REQ_x\|^{2}=\frac{1}{m}\sum_{b \in M}a'^{d_H(b,x)}(1-a')^{ n-d_H(b,x)}.
\end{equation}
where $m$ is the number of patterns for the learning, $b$ is an item of the set
$M$ of patterns, and we choose the case where $a'=a_{b}$ is the same for all the
patterns.

\end{itemize}

It is noteworthy that using a straight forward approach to classification and
employing Grover's search, Ventura have found that the exclusion method exhibit
the lowest overall probability of irrelevant classification compared to
inclusion method \cite{ventura.jcis02}. This explains why the exclusion method
is
use in the Algorithm \ref{alg:requetm}.

\section{Simulations and results}

\begin{figure}[ptbh]
\centering
\begin{minipage}[c]{.55\linewidth}
\centering
 \includegraphics[scale=.6,keepaspectratio=true]{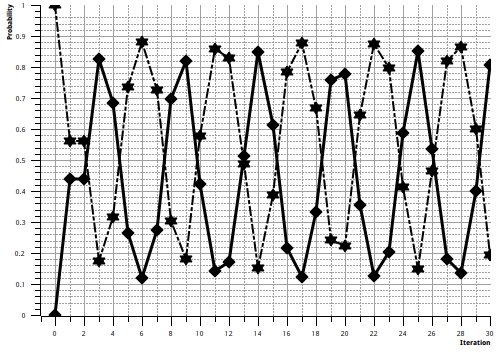}
\vspace*{13pt}
 \caption{Probability evolution with the number of iterations for
\textbf{C1}-algorithm for 7-qubits memory patterns. The solid line represents
the probability $P_c$ of a correct recognition and the dotted line the
probability $P_w$ of an incorrect recognition.}
 \label{fig:ImprovEzhovC1}
\end{minipage} \hfill
\begin{minipage}[c]{.33\linewidth}
\centering
 \includegraphics[scale=.6]{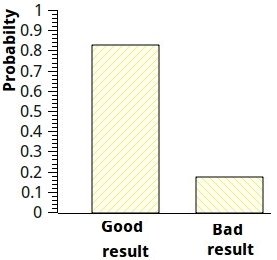}
\vspace*{13pt}
 \caption{Probability of correct recognition for a set of two example patterns
for \textbf{C1}-algorithm. }
 \label{fig:ProbaEzhovC1}
\end{minipage}
\end{figure}

\begin{figure}[ptbh]
\centering
\begin{minipage}[c]{.55\linewidth}
\centering
 \includegraphics[scale=.55,keepaspectratio=true]{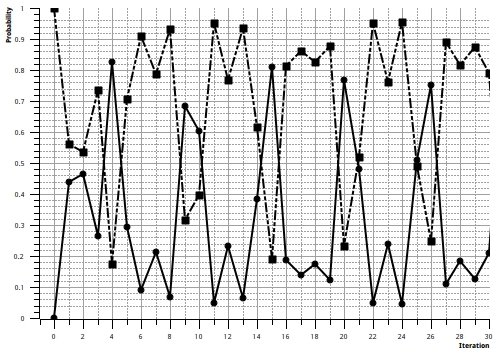}
\vspace*{13pt}
 \caption{Probability evolution with the number of iterations for
\textbf{C2}-algorithm for 7-qubits memory patterns. The solid line represents
the probability $P_c$ of a correct recognition and the dotted line the
probability $P_w$ of an incorrect recognition.}
 \label{fig:ImprovEzhovC2}
\end{minipage} \hfill
\begin{minipage}[c]{.38\linewidth}
\centering
 \includegraphics[scale=.5]{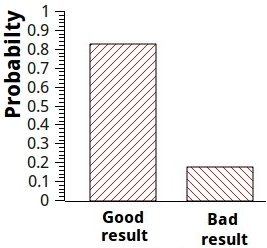}
\vspace*{13pt}
 \caption{Probability of correct recognition for a set of two example patterns
for \textbf{C2}-algorithm.}
 \label{fig:ProbaEzhovC2}
\end{minipage}
\end{figure}

Consider the data of Example \ref{exa:exam1}, the evolution of probabilities
with the number of iterations are plot in Figures \ref{fig:ImprovEzhovC1} and
\ref{fig:ImprovEzhovC2} for \textbf{C1}-algorithm and \textbf{C1}-algorithm
respectively.

For \textbf{C1}-algorithm,
\begin{equation}
 \mathcal{I}_M=\mathbb{I}-(\ket{2}\bra{2}+\ket{4}\bra{4}),\,\Lambda=25\,(T=11).
\end{equation}
Therefore, the steps \ref{etap6} and \ref{etap7} of Algorithm \ref{alg:requetm}
are repeated $23$ times. At the end of the algorithm, the register is in the
state
\begin{equation}
 \begin{split}
\ket{\psi^{25}} & =-0.137\ket{0}+0.0231\ket{1}-0.876\ket{2}+0.301\ket{3}
-0.292\ket{4}-0.137\ket{5}\\ & -0.137\ket{6}+0.0231\ket{7}.
\end{split}
\end{equation}
The probability to retrieve the memory states, $\ket{2}$ and $\ket{4}$, is
$0.86266^2+0.28755^2=85.23\%$. The ratio
\begin{equation}
\frac{P_c}{P_w}=\frac{0.852253}{0.1477}=5.77,
\end{equation}
shows how higher recognition efficiency of \textbf{C1}-algorithm compared to
Ezhov's model.

For \textbf{C2}-algorithm, by choosing $a'=0.1$, one find
\begin{equation}
 \begin{split}
\ket{REQ}=0.285\ket{0}+0.095\ket{1}+0.607\ket{2}+0.202\ket{3}+0.607\ket{4}\\+
0.202\ket{5} +0.285\ket{6} +0.095\ket{7},
 \end{split}
  \end{equation}
\begin{equation}
 \begin{split}
 \mathcal{I}_M & =\mathbb{I}-(0.285\ket{0}+0.095\ket{1}+0.607\ket{2}
+0.202\ket{3}+0.607\ket{4}+0.202\ket{5}\\ & +0.285\ket{6} +0.095\ket{7})
(0.285\bra{0}+0.095\bra{1}+0.607\bra{2}+0.202\bra{3}\\
& +0.607\bra {4}+0.202\bra{5} +0.285\bra{6} +0.095\bra{7}).
 \end{split}
  \end{equation}
and $\Lambda=4$. Therefore, the steps \ref{etap6} and \ref{etap7} of Algorithm
\ref{alg:requetm} are repeated $2$ times. At the end of the algorithm, the
register is in the state
\begin{equation}
 \begin{split}
 \ket{\psi^{4}} &=-0.107\ket{0}-0.024\ket{1}+0.772\ket{2}+0.358\ket{3}+0.477\ket{
4}+0.152\ket{5}\\& -0.107\ket{6}-0.024\ket{7}.
 \end{split}
\end{equation}
The probability of correct recognition of memory states $\ket{2}$ and $\ket{4}$
is $0.772^2+0.477^2=82.69\%$, which is fairly lower than that of
\textbf{C1}-algorithm.

The Table \ref{tab:3qubit} which gives the summary of the relevant parameters of
Ezhov's, \textbf{C1} and \textbf{C2} methods shows that despite the
\textbf{C1}-algorithm is slower than \textbf{C2}-algorithm it leads to a better
recognition efficiency of memory patterns. There is a significant gap between
the recognition efficiency of \textbf{C1} and \textbf{C2} algorithms and that of
Ezhov's model.

\vspace*{4pt}
\begin{table}[htp]
\caption{Relevant parameters of Ezhov's, \textbf{C1} and \textbf{C2} methods
for 3-qubits memory data of Example \protect\ref{exa:exam1}.}
\centering
\begin{tabular}{lrr}\hline
Method & $\Lambda$ & $P_c/P_w$\\\hline
Ezhov's & 4 & 1.08\\
\textbf{C1} & 25 & 5.77\\
\textbf{C2} & 4 & 4.67\\\hline
\end{tabular}
\label{tab:3qubit}
\end{table}

\begin{figure}[!ht]
 \centering
 \includegraphics[scale=.4]{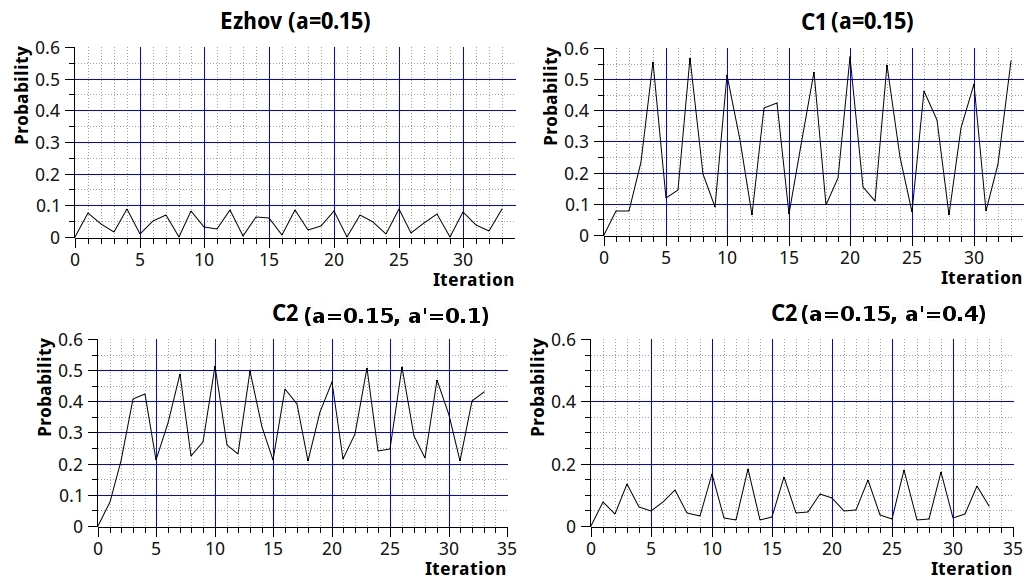}
\vspace*{13pt}
 \caption{Probability evolution with the number of iterations for 7-qubits
memory patterns. The arbitrary value that regulates the width of the
distribution is $a=0.15$.}
 \label{fig:ImprovEvol7Qubits15}
\end{figure}

\begin{figure}[!ht]
 \centering
 \includegraphics[scale=.4]{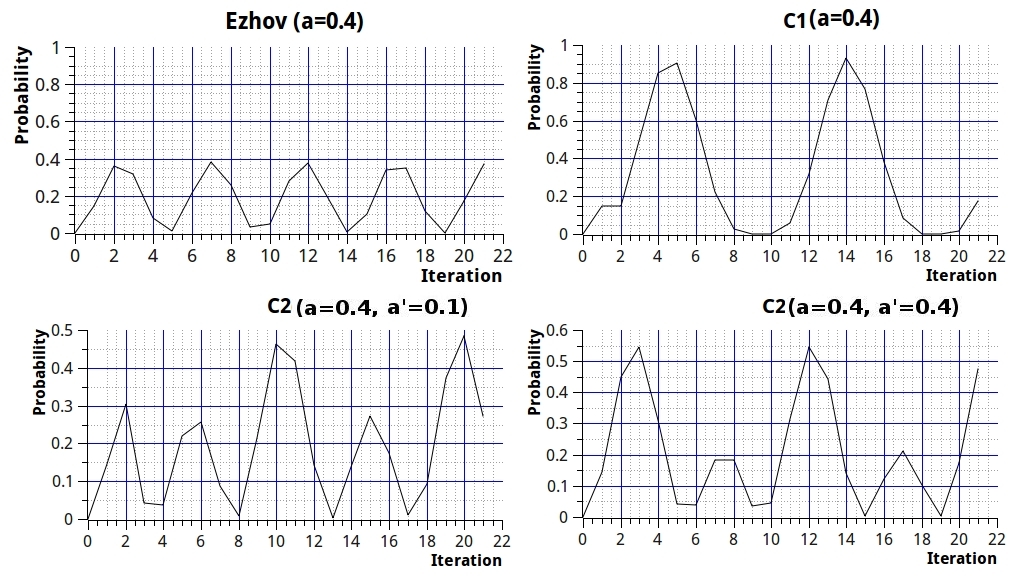}
\vspace*{13pt}
 \caption{Probability evolution with the number of iterations for 7-qubits
memory patterns. The arbitrary value that regulates the width of the
distribution is $a=0.40$.}
 \label{fig:ImprovEvol7Qubits40}
\end{figure}

For a better comparison of \textbf{C1} and \textbf{C2} algorithms, consider a
7-qubits memory where the patterns for the learning are states $\ket{23}$,
$\ket{59}$, $\ket{61}$, and $\ket{110}$ and the distribution query centered on
$\ket{60}$. Figures \ref{fig:ImprovEvol7Qubits15} and
\ref{fig:ImprovEvol7Qubits40} show the corresponding probabilities of observing
the correct recognition upon system measurement versus the number of iterations
of Grover's search for $a=0.15$ and $a=0.40$ respectively.

\vspace*{4pt}
\begin{table}[htp]
\caption{Relevant parameters of Ezhov's, \textbf{C1} and \textbf{C2} methods
for 7-qubits memory data.}
 \centering
\begin{tabular}{lrrrr}\hline
Method & $a$ & $a'$ & $\Lambda$ & $P_c$\\\hline
Ezhov's & $0.15$ & - & $32$ & $<10\%$\\
\textbf{C1} & $0.15$ & - & $20$ & $57.04\%$\\
\textbf{C2} & $0.15$ &$0.10$ & $10$ & $51.45\%$\\
\textbf{C2} & $0.15$ & $0.40$ & $13$ & $18.35\%$\\\hline
Ezhov's & $0.40$ & - & $21$ & $<40\%$\\
\textbf{C1} & $0.40$ & - & $14$ & $93.22\%$\\
\textbf{C2} & $0.40$ &$0.10$ & $20$ & $48.37\%$\\
\textbf{C2} & $0.40$ & $0.40$ & $12$ & $54.70\%$\\\hline
 \end{tabular}
\label{tab:7qubit}
\end{table}

It can be seen, as summarize in the Table \ref{tab:7qubit}, that

\begin{itemize}
\item the smaller the arbitrary value that regulates the width of the
distribution $a$, for the three methods
\begin{itemize}
\item the larger number of iterations $\Lambda$;
\item the lower the recognition efficiency;
\end{itemize}

\item for a giving value of $a$ in the \textbf{C2}-algorithm, the closer $a'$
\begin{itemize}
\item the lower number of iterations $\Lambda$;
\item the larger the recognition efficiency;
\end{itemize}

\item for any value value of $a\in[0,1/2]$,
\begin{itemize}
 \item the best recognition efficiency of memory patterns is given by
\textbf{C1}-algorithm and the poorest by the standard Ezhov's model;
\item \textbf{C2}-algorithm seem to be faster than \textbf{C1}-algorithm.
\end{itemize}

\end{itemize}

\section{Conclusion}
\label{esc:concl}

In this paper we have proposed a multi-pattern recognition method with a good
rate of success based on an improved quantum associative algorithm with
distributed query for any rate of the number of the recognition pattern on the
total patterns. We have introduced an operator $\mathcal{I}_{M}$ which acts as
the oracle operator by considering two cases: the case \textbf{C1} where
$\mathcal{I}_{M}$ invert only the phase of the memory patterns states as in the
Ventura's model; the case \textbf{C2} where $\mathcal{I}_{M}$ invert the
probability amplitudes of all the states over the average amplitude of the query
state centered on the $m$ patterns of the learning. These improvements appeared
as factors that increased constructive interferences so that the number of
iterations is considerably reduced. Simulation results favour the
\textbf{C1}-algorithm despite his high number of iterations compare to the
\textbf{C2}-algorithm. Work is in progress to investigate other possibilities of
improvement.



\begin{thebibliography}{10}
\bibitem{perus-bischof-03}
Perus M., Bischof H.:
\newblock {Quantum-wave pattern recognition: From simulations towards
implementation},
\newblock Proceedings 7th Joint Conf. Information Sciences 2003, pp. 1536-1539
(2003)

\bibitem{PerusBischof:2003}
Perus M., Bischof H.:
\newblock {A neural-network-like quantum information processing system},
\newblock quant-ph/0305072 (2003).

\bibitem{perus.2004-43}
Perus M, Bischof H., Loo C.K.:
\newblock {Quantum-implemented selective reconstruction of high-resolution
images},
\newblock Appl. Opt., 43, pp. 6134, 2004; quant-ph/0401016 (2004).

\bibitem{brickman.2005}
{Brickman, K.-A., Haljan, P.~C., Lee, P.~J., Acton, M., Deslauriers, L.,
  Monroe, C.:}
\newblock Implementation of grover's quantum search algorithm in a scalable
  system.
\newblock {Phys. Rev. A 72\/}, 050306 (2005).

\bibitem{ezhov.is00}
{Ezhov, A., Nifanova, A., Ventura, D.:}
\newblock Distributed queries for quantum associative memory.
\newblock {Information Sciences 3-4\/}, 271--293 (2000).

\bibitem{grover.1998-80}
{Grover, L.:}
\newblock Quantum computers can search rapidly by using almost any
  transformation.
\newblock {Phys. Rev. Lett. 80\/}, 4329--4332 (1998).

\bibitem{li2007phase}
{Li, P., Li, S.:}
\newblock Phase matching in grover's algorithm.
\newblock {Physics Letters A 366}, 1-2, 42--46 (2007).

\bibitem{Trugenberger:02}
{Trugenberger, C.:}
\newblock Quantum pattern recognition.
\newblock {Phys. Rev. Lett. 89\/}, 277903 (2002).

\bibitem{ventura.ijcnn01}
{Ventura, D.:}
\newblock On the utility of entanglement in quantum neural computing.
\newblock In { Proceedings of the International Joint Conference on Neural
  Networks\/}, pp.~1565--1570 (2001).

\bibitem{ventura.jcis02}
{Ventura, D.:}
\newblock Pattern classification using a quantum system.
\newblock In {Proceedings of the Joint Conference on Information
  Sciences\/}, pp.~537--640 (2002).

\bibitem{ventura.fopl99}
{Ventura, D., Martinez, T.:}
\newblock Initializing the amplitude distribution of a quantum state.
\newblock {Foundations of Physics Letters 6\/}, 547--559 (1999).

\bibitem{ventura.is00}
{Ventura, D., Martinez, T:.}
\newblock Quantum associat:ive memory.
\newblock {Information Sciences 1-4\/}, 273--296 (2000).

\bibitem{ZhouDing:2008}
{Zhou, R., Ding, Q.:}
\newblock Quantum pattern recognition with probability of 100\%.
\newblock {International Journal of Theoretical Physics 47}, 5,
  1278--1285 (2008).
 \end{thebibliography}

\appendix
\section{Learning algorithms}

\begin{algorithm}
\caption{Learning the set $M$ by initializing the amplitude distribution of a
quantum state (Ref.~\protect\cite{ventura.fopl99})}
\label{alg:apprend}
\begin{algorithmic}[1]
\STATE $\ket{\psi}=\ket{x_{1}\dots x_{n},g_{1}\dots
g_{n-1},c_{1}c_{2}}\equiv\ket{0_1\dots0_n,0_1\dots0_{n-1},00}$;
\COMMENT{Initialization}
\FOR{$m\geq p\geq 1$}
\FOR{$1\leq j\leq n$}
\IF{$z_{pj}\neq z_{(p+1)j}$}
\STATE $\ket{\psi}=\mathtt{CNOT}_{(c_{2}x_{j})}^{0}\ket{\psi}$;
\ENDIF
\ENDFOR
\STATE $\ket{\psi}=\mathtt{CNOT}_{(c_{2}c_{1})}^{0}\ket{\psi}$;
\STATE $\ket{\psi}=\mathtt{F}^{p}\ket{\psi}$;
\STATE $\ket{\psi}=\mathtt{F}^{z_{p1}z_{p2}}_{x_{1}x_{2}g_{1}}\ket{\psi}$;
\FOR{$3\leq k\leq n$}
\STATE $\ket{\psi}=\mathtt{F}^{z_{pk1}}_{x_{k}g_{k-2}g_{k-1}}\ket{\psi}$;
\ENDFOR
\STATE $\ket{\psi}=\mathtt{CNOT}_{(g_{n-1}c_{1})}^{1}\ket{\psi};$
\FOR{$n\geq k\geq 3$}
\STATE $\ket{\psi}=\mathtt{F}^{z_{pk1}}_{x_{k}g_{k-2}g_{k-1}}\ket{\psi}$;
\ENDFOR
\STATE $\ket{\psi}=\mathtt{F}^{z_{p1}z_{p2}}_{x_{1}x_{2}g_{1}}\ket{\psi}$;
\ENDFOR
\STATE $\ket{\psi}=\mathtt{NO}T_{c_{2}}\ket{\psi}$;
\STATE Observe the system.
\end{algorithmic}
\end{algorithm}

The algorithm \ref{alg:apprend} use the following gates:
\begin{itemize}
\item  The hermitian 1-qubit \texttt{NOT} gate
\begin{equation}
 \mathtt{NOT}=\ket{0}\bra{1}+\ket{1}\bra{0} =\begin{pmatrix}
 0 & 1\\
 1 & 0
\end{pmatrix}.
\end{equation}

\item Two forms of the hermitian 2-qubit controlled gate $\mathtt{CNOT}$
\begin{equation}
 \mathtt{CNOT}^{0}=\mathtt{diag}(\mathtt{NOT},\mathbb{I})
\text{ and }\mathtt{CNOT}^{1}=\mathtt{diag}(\mathbb{I},\mathtt{NOT}).
\end{equation}
\item Four forms of of the hermitian 3-qubit Fredkin's gate $\mathtt{F}$ (a
controlled-controlled NOT gate)
\begin{align}
\mathtt{F}^{00}
& =\mathtt{diag}(\mathtt{NOT},\mathbb{I},\mathbb{I},\mathbb{I}),\,
\mathtt{F}^{01}=\mathtt{diag}(\mathbb{I},\mathtt{NOT},\mathbb{I},\mathbb{I}),\\
\mathtt{F}^{10} &
=\mathtt{diag}(\mathbb{I},\mathbb{I},\mathtt{NOT},\mathbb{I}),\,\mathtt{F}^{11}
=\mathtt{diag}(\mathbb{I},\mathbb{I},\mathbb{I},\mathtt{NOT}).
\end{align}
\item And $z_{pj}$ the values of different qubits of patterns, where we take
$z_{(m+1)j}=0^{\otimes m}$.
\end{itemize}

\end{document}